\documentclass[a4paper,fleqn,usenatbib]{mnras}

\usepackage[T1]{fontenc}
\usepackage{ae,aecompl}
\usepackage{graphicx}	
\usepackage{amsmath}	
\usepackage{amssymb}	

\usepackage{cite}

\title[Radial eigenmodes of levitating atmospheres]{Radial modes of levitating atmospheres around Eddington-luminosity neutron stars}

\author[D. A. Bollimpalli and W. Klu\'{z}niak]{
D. A. Bollimpalli,$^{}$\thanks{E-mail: \href{mailto:deepika@camk.edu.pl}{deepika@camk.edu.pl}}
W. Klu\'{z}niak$^{}$\thanks{E-mail: \href{mailto:wlodek@camk.edu.pl}{wlodek@camk.edu.pl}}
\\
$^{}$Nicolaus Copernicus Astronomical Center, ul. Bartycka 18, PL 00-716 Warsaw, Poland\\
$^{}$KITP, University of California Santa Barbara, CA 93106, USA\\
}

\date{Accepted XXX. Received YYY; in original form ZZZ}

\pubyear{2016}

\begin{document}
\label{firstpage}
\pagerange{\pageref{firstpage}--\pageref{lastpage}}
\maketitle

\begin{abstract}
We consider an optically thin radiation-supported levitating atmosphere suspended well above the stellar surface, as discussed recently in the Schwarzschild metric for a star of luminosity close to the Eddington value. Assuming the atmosphere to be geometrically thin and polytropic, we investigate the eigenmodes and calculate the frequencies of the oscillations of the atmosphere in Newtonian formalism. The ratio of the two lowest eigenfrequencies is $\sqrt{\gamma+1}$, i.e., it only depends on the adiabatic index.
\end{abstract}


\section{Introduction}

Recent observations have mapped the existence of neutron stars with highly super-Eddington luminosities in ultra-luminous X-ray source (ULX) systems NGC 7793 P13, NGC 5907, NuSTAR J09551+6940.8 \citep{bachetti, israel2016, israel2017}. It has long been known that neutron stars in X-ray binaries, A0538-66, SMC X-1 and GRO J1744--28 are unusually luminous \citep{1982nat, groj, engel}. Also, studies of X-ray bursts report that, during very energetic Type I X-ray bursts, the X-ray luminosity is close to the Eddington limit \citep{book26}. For neutron stars with such high luminosities, it is possible that close to the star's surface, radiative forces become dominant over gravity. Under such conditions a ``levitating atmosphere'' may be formed, the properties of which were derived in the spherically symmetric case by  \citet{maciek2015,maciek2016}, and had previously been discussed for the approximately plane- parallel case in the  context of near Eddington luminosity accretion discs by \citet{fukue1996}, who termed them `photon floaters'. An additional argument of the observational relevance for these levitating atmospheres has been afforded in a new study, in which they have been shown to affect the propagation of light rays from the central source, thereby affecting the observed appearance of the central object \citep{adam2017}.

X-ray bursters exhibit  high frequency (300 -- 600 Hz) `burst oscillations' \citep{stro2001}, which may be related to the Eddington-luminosity of the sources. In bright X-ray bursts, neutron star often reaches near Eddington-limit luminosities during which the dominant radiation pressure drives the photospheric layers off the neutron star; this phenomenon is termed as photospheric radius expansion (PRE). Oscillations in the case of PRE bursts are observed during the rise and decay phase of the burst \citep{muno2002}. However, bursts without PRE also show evidence for oscillations~\citep{van2001}. During the decay phase of the burst, frequency is observed to increase with time \citep{stro2006}. While the source of these oscillations is still open for debate, here we try to understand if the radial oscillations of the levitating atmospheres could be relevant to such phenomena.

In this paper, we consider a radial perturbation of a spherically symmetric, optically thin levitating atmosphere. The goal is to analytically investigate the eigenmodes of the oscillations of the atmosphere. 
\citet{abarca2016} have found the fundamental radial mode of oscillations of an optically thin levitating atmosphere. They have shown that including the radiation drag stabilizes the atmosphere against the radial perturbations that could excite the fundamental radial mode.
Here, we would like to investigate all possible modes of radial oscillations for this given system.
 Assuming the atmosphere to be geometrically thin, we neglect the relativistic corrections and perform linear perturbation analysis in the Newtonian approximation. We then determine all the possible radial eigenmode solutions of the system. Our calculations are performed by perturbing the  levitating atmospheres of \citet{maciek2015}, which have the form of spherical shells, but the calculations should be equally valid for the plane parallel case of the photon floaters of \citet{fukue1996}, and have some application as well to the eigenmodes of other accretion structures with a reflection symmetry, such as the slender tori of \citet{blaes2006}.

Optically thin levitating atmospheres are especially likely to survive in the pulsating ULXs, which can be understood as slowly rotating neutron stars, a special case of extremely luminous X-ray pulsars \citep{bachetti}, in which the super-Eddington accretion flow near the poles is confined by the strong stellar magnetic field \citep{Basko1976, Mushtukov2017}.
  
A study of optically thick atmospheres would have been more directly relevant to X-ray burst oscillations, however we are hampered at present by the lack of a suitable analytic solution for the background equilibrium solution. The optically thick super-Eddington atmosphere solution  \citep{maciek2016} requires a proper treatment of the radiative transfer that has only been achieved numerically.

\section{Levitating atmospheres}

\citet{marek1990} have studied the fully general relativistic radial motion of test particles in combined gravitational and radiation fields around a spherical, compact star radiating isotropically. They have shown that under these circumstances, a spherical shell of matter may form around the neutron star whose location varies with luminosity from the surface of the star to infinity. This surface is defined by the radius where the gravitational forces balance the radiative forces due to stellar luminosity. Unlike in the Newtonian theory, where both the gravity and radiative flux fall off as $1/r^2$, in general relativity, the radiative flux has a stronger radial dependence than the gravitational acceleration. Therefore, radiative forces may dominate over gravitational pull close to the surface of the compact star, while the gravitational forces take over at larger radii. A similar situation may obtain above nearly Eddington-luminosity accretion disc, as discussed in several papers \citep{fukue1996,2003PASJ...55.1115K,2017PASJ...69...56I}.

For a star radiating at super-Eddington luminosities, gravitational forces will balance the radiation forces on a sphere at a particular radius from the star. It has been referred to as the `Eddington capture sphere' (ECS) \citep{stahl2012}.
A test particle moving with a reasonable velocity would bind to this sphere without falling on to the stellar surface also if initially in non-radial motion as the Poynting-Robertson-radiation drag removes the angular momentum of the particle reducing it to zero \citep{bini2009, oh2010, stahl2013}. The captured particles levitate above the stellar surface, as every point on ECS corresponds to an equilibrium state that is stable in radial direction \citep{marek1990} and neutral in directions tangent to the sphere \citep{stahl2012}. These properties of ECS result in an atmosphere in hydrostatic equilibrium, which is detached from the star and levitating above its surface \citep{maciekp}. In the case of an optically thin atmosphere, analytical solutions show that both the pressure and density fall steeply on either sides of the ECS, while the peak value is attained at the ECS \citep{maciek2015}. The term ``levitation'' was first introduced by \citet{is1999} in the context of the luminous equatorial accretion belt (boundary layer) formed at the interface of the neutron star surface and the accretion disc; the effective gravity in the boundary layer is further reduced by the rapid azimuthal motion of the fluid, so the balance considered by these authors was between gravity and three forces: pressure gradient, centrifugal and radiation forces. Our simplified treatment neglects rotation.

\section{Hydrostatic Equilibrium Configuration}

In the Schwarzschild metric around a neutron star of mass $M$, a local observer at a certain radius $r$ measures the redshifted luminosity of the star, $L(r)$ as
\begin{equation}
L(r)= L_{\infty}\left(1-\dfrac{2r_\mathrm{g}}{r} \right)^{-1}.
\label{lumin}
\end{equation} 
where $L_{\infty}$ is the luminosity of the star measured at infinity and $r_{\mathrm{g}}= GM/c^2$ is the gravitational radius. For a compact star of radius $R_{*}$ and luminosity $L_{*}$, we have $L_{\infty} = L_{*}\left(1-\dfrac{2r_{\mathrm{g}}}{R_{*}} \right)$. For convenience, we shall use the following parametrized luminosity throughout our calculations,
\begin{equation}
\lambda = L_{\infty}/L_{\mathrm{Edd}}, 
\label{lambda}
\end{equation}
where $L_{\mathrm{Edd}}$ is the Eddington luminosity.

We now consider a static atmosphere levitating above the surface of the neutron star radiating at super-Eddington luminosities. The equation of hydrostatic equilibrium for an optically thin atmosphere under relativistic formalism has been derived previously \citep{maciek2015}:
\begin{equation}
\dfrac{1}{\rho}\dfrac{dp}{dr}=-\dfrac{GM}{r^2\left(1-r_{\mathrm{g}}/r\right)}\left[1-\lambda\left(1-\dfrac{2r_{\mathrm{g}}}{r}\right)^{-1/2}\right].
\label{eqbrm}
\end{equation}

The first term on the right-hand side comes from the gravitational force, while the second term is related to the radiation force in an optically thin fluid. In deriving the above Eq. (\ref{eqbrm}), it has been assumed that the atmospheric temperatures are non-relativistic, i.e., of the order of few keV. At such temperatures, one can neglect the internal energy and pressure of the fluid in comparison to the rest mass energy density. 

From the above equation, it follows that the gravitational force and the force due to radiation balance each other at radius 
\begin{equation}
R_{\mathrm{ECS}} \equiv \dfrac{2r_{\mathrm{g}}}{(1-\lambda^2)},
\label{recs}
\end{equation}
and the pressure gradient at this radius becomes zero, corresponding to the maximum pressure at $R_{\mathrm{ECS}}$.

We consider barotropic fluid, $p = p(\rho)$, and in this case we assume it to be a polytrope. Therefore
\begin{equation}
p = K \rho^\mathrm{1+1/n},
\label{eqstate}
\end{equation}
where $K$ is the polytropic constant and $n$ is the polytropic index. 

We may cast Eq. (\ref{eqbrm}) in terms of an effective potential $\Phi(r)$, in such a way that upon its integration using Eq. (\ref{eqstate}), it results in the Bernoulli equation,
\begin{equation}
(n+1)\dfrac{p}{\rho} + \Phi(r) = \mathrm{const}.
\label{e4}
\end{equation}

We may evaluate this constant on the equilibrium surface, $r_0 \equiv R_{\mathrm{ECS}}$, to get 
\begin{equation}
 (n+1)\dfrac{p}{\rho} +\Phi(r) = 
(n+1)\dfrac{p_0}{\rho_0} + \Phi(r_0),
\label{e5}
\end{equation}
where subscript zero denotes the quantities evaluated at the ECS, $r=r_0$. Following simple algebraic steps, one could write
\begin{eqnarray}
\dfrac{p}{\rho}&=& \dfrac{p_0}{\rho_0} f,
\label{e6}
\end{eqnarray}
where we define the function $f(r)$ as 
\begin{eqnarray}
f(r) &=& 1 - \dfrac{1}{n c_{\mathrm{s,0}}^2}\left[\Phi(r) - \Phi(r_0)\right] 
\label{e7}
\end{eqnarray}
and $c_{\mathrm{s}}^2 = \dfrac{(n+1)}{n}\dfrac{p}{\rho}$ is the square of the sound speed. 

We now evaluate this function $f$ in the vicinity of $r_0$ by the second-order Taylor expansion of the effective potential, which is valid for thin atmospheres. We have
\begin{eqnarray}
\Phi (r)- \Phi(r_0) &=&  \dfrac{(1-\lambda^2)^4c^2}{32\lambda^4}  \dfrac{r_0^2}{r_{\mathrm{g}}^2}x^2 ,
\label{e11}
\end{eqnarray}
where, we have introduced a new coordinate $x =(r-r_0)/r_0$, centred at $r_0$. This will immediately give us
\begin{eqnarray}
f &=& 1- \dfrac{x^2}{\beta ^2}\omega_{r,0}^2,
\label{e15}
\end{eqnarray}
where we define $\beta^2 = 2nc_{\mathrm{s,0}}^2/r_0^2$ and
\begin{eqnarray}
  \omega_{r,0}^2 = \dfrac{(1-\lambda^2)^4}{16\lambda^4}\dfrac{c^2}{r_{\mathrm{g}}^2}
 \label{fundament}
\end{eqnarray}
for our convenience. We shall later show that $\omega_{r,0}$ is indeed the frequency of the fundamental mode. The boundaries of the atmosphere are defined at $f=0$ as both the pressure and density vanish at this point. Following this, the thickness of the atmosphere would be
\begin{eqnarray}
\Delta r = 2\beta r_0/\omega_{r,0} = 4\sqrt{2n}\left(\dfrac{c_{\mathrm{s},0}}{c}\right)\left(\dfrac{\lambda^2}{1-\lambda^2}\right)
\label{thick}
\end{eqnarray}

\begin{figure}
\begin{center}
\includegraphics[width=\columnwidth]{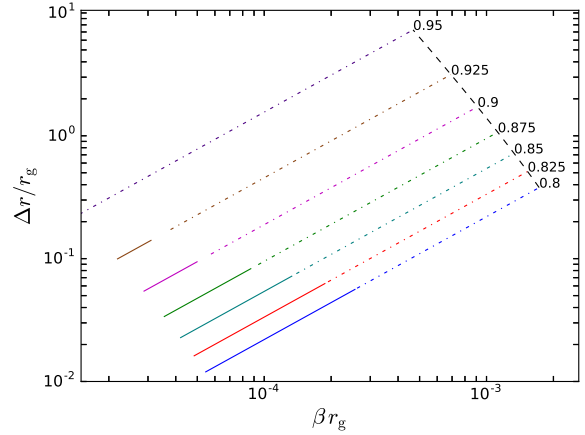}
\caption{Atmosphere thickness, $\Delta r$ dependence on $\beta$ and $\lambda$ parameters. Both the axes are logarithmically scaled. Set of solid and dash-dot lines labelled with values of $\lambda$, represent the linear relation between thickness of the atmosphere and $\beta$. The solid lines mark the regime where the second order approximation holds good. The dashed line is plotted for a particular temperature value $10^8~\mathrm{K}$ that shows that atmosphere gets thinner as $\lambda$ decreases. We considered $\mu = 1/2$ for this plot.}
\label{fig:mesh3}
\end{center}
\end{figure}

In Fig. \ref{fig:mesh3}, we explore the dependence of the atmosphere thickness, $\Delta r$ on $\beta$ and $\lambda$. Each solid and dash-dot line, representing a particular value of $\lambda$, shows that $\Delta r$ increases with $\beta$. Specifically, solid lines represent the regime where the second order approximation may hold good, i.e., $x \ll 1$ (we simply chose $x = 0.005$ for the plot). Therefore, for a star with a given luminosity, $\lambda$ remains constant, and the extent to which the atmosphere extends is determined by $\beta$ and in the slender limit $\beta \rightarrow 0$.

As $\beta$ implicitly depends on temperature at $r_0$, in Fig. \ref{fig:mesh3} we considered the temperature range within $[10^5, 10^8]~\mathrm{K}$ and we take the mean molecular weight of the fluid, $\mu$ to be $1/2$. The dashed line in Fig. \ref{fig:mesh3} is plotted to give a glimpse of how $\Delta r$ varies with $\lambda$. As shown in the figure, the slenderness of the atmosphere increases as $\lambda$ decreases. Here the dashed line is plotted for the temperature value, $10^8~\mathrm{K}$, but the trend would remain the same for other values of temperature.
 
Based on Eq. (\ref{e6}) and the equation of state, Eq. (\ref{eqstate}), it is convenient to scale the pressure and density in equilibrium as follows:
\begin{eqnarray}
p &=& p_0 f^{n+1}, \nonumber\\
\rho &=& \rho_0 f^n,
\label{prho}
\end{eqnarray}
which clearly ensures that the pressure and density drop to zero at the boundaries while retaining their maximum value at the centre. The density gradient is always in the direction of the effective gravity (taking into account radiation forces as well), and so is Rayleigh-Taylor stable.

In passing, we note an interesting property of the numerical solution for the optically thick atmosphere, which may make our analysis an interesting limiting case of future (perhaps numerical) calculations of the oscillations of the optically thick levitating atmosphere. The optically thin edges of the optically thick atmosphere coincide with this analytically derived optically thin solution, while the radial density has a universal profile (independent of the mass of the atmospheric shell) in the optically thick regions, satisfying over the bulk of the shell the condition of density decreasing with radial distance from the source, i.e., increasing in the direction of the effective force of gravity \citep{maciekp}.

\section{Elementary discussion of radial oscillations}

To develop intuitions relevant to fluid motion in a gravitational field of magnitude increasing with height we present an elementary discussion in a Newtonian formalism of the lowest frequency radial modes of a thin levitating atmosphere in the plane-parallel approximation.

Choose the $z$ axis in the vertical direction, pointing either towards or away from the neutron star, and consider a fluid atmosphere in a gravitational field directed towards the $z=0$ plane: 
$\vec g(z) = -\hat z g(z)$, with $g(z)=\omega^2_\perp z$. In hydrostatic equilibrium, the surface of maximum pressure coincides with the $z=0$ plane, and the atmosphere is symmetric with respect to reflection in that plane.

Let us consider the lowest-frequency modes of the atmosphere, one with a symmetric displacement function $z \mapsto z+\xi(z,t)$, $\xi(-z,t)=\xi(z,t)$, and no nodes (the ``fundamental"), and one with antisymmetric displacements, $\xi(-z,t)=-\xi(z,t)$, and one node (the ``first overtone"). As we will see, the fundamental mode corresponds to rigid motion, with the plane of non-varying maximum pressure moving harmonically up and down with respect to the $z=0$ plane. The first overtone is a breathing-like motion characterized by reflection symmetry in the $z=0$ plane. In general, as the atmosphere oscillates, its pressure is a function of position and time $p(z,t)$.

To derive the eigenfrequencies, consider --- without any loss of generality --- the part of the atmosphere that is initially above the horizontal plane at $z=z_0>0$ plane. For a polytropic atmosphere the density goes to zero at the upper boundary initially at, say, $z=H>0$ (and another one at $z=-H$). Thus, we are focusing on atmospheric slab of matter confined by the upper boundary (plane) at $H+ \xi(H,t)$ and an imagined lower boundary at $z_0+\xi(z_0,t)$. The conserved mass per unit area in this slab is
\begin{equation}
m = \int \mathrm{d}m = \int^H_{z_0} \rho(z) \mathrm{d}z.
\label{mass2}
\end{equation}
The slab suffers two external forces, a body force owing to gravity, and a surface force at the lower boundary at $z_0+\xi$. The latter force per unit area is just the pressure $p = p_0 + \Delta p(t)$, with $p_0$ the equilibrium pressure at $z=z_0$, and $\Delta p$ the Lagrangian change in pressure: $\Delta p(t) = p(z_0+\xi_0,t) - p_0$, with $\xi_0 \equiv \xi(z_0,t)$. The body force is just the weight of the slab, per unit area
\begin{equation}
w(t) = \int g\, \mathrm{d}m = \int^H_{z_0} g(z+\xi) \rho(z) \mathrm{d}z.
\label{centermass}
\end{equation}
We can now specify to the case at hand, $w_0 = m \omega^2_\perp z_{\mathrm CM}$ with $z_{\mathrm CM} = (\int z\,\mathrm{d}m)/m$,
\begin{equation}
w(t)= w_0 + \omega^2_\perp \int^H_{z_0} \xi(z,t) \rho(z) \mathrm{d}z.
\label{weight2}
\end{equation}
The equation of motion for the displacement of the centre of mass, $\xi_{\mathrm CM}(t) \equiv \xi(z_{\mathrm CM},t)$, is
\begin{equation}
m \partial^2 \xi_{\mathrm CM}/\partial t^2 = p - w.
\end{equation}

For a uniform displacement, $\xi(z,t) = \xi_{\mathrm CM}(t)$, i.e., for a rigid motion, there are no nodes, and we obtain the ``fundamental" frequency $\omega_1 = \omega_\perp$ from the simple harmonic equation $\partial^2 \xi_{\mathrm CM}/\partial t^2 = - \omega^2_\perp \xi_{\mathrm CM}$, following from $\partial \xi/ \partial z =0$ that implies $p = p_0 =w_0$, 
with $p-w= - m \omega^2_\perp \xi_{\mathrm CM}$ by Eqs. (\ref{mass2}) and (\ref{weight2}).

For the breathing mode, with the (easy to derive) ansatz $\xi = z \xi_{\mathrm CM}/z_{\mathrm CM}$, we have $\Delta p = - \gamma p_0 \xi_{\mathrm CM}/z_{\mathrm CM} = - \gamma m \omega^2_\perp \xi_{\mathrm CM}$, and $w(t) = w_0 + m \omega^2_\perp \xi_{\mathrm CM}(t)$. Thus
$$
\partial^2 \xi_{\mathrm CM}/\partial t^2 + (\gamma +1) \omega^2_\perp \xi_{\mathrm CM} = 0,
$$
and the eigenfrequency of the breathing mode is $\omega_2 = \sqrt{1+\gamma} \omega_1$.
The same ratio of $\sqrt{1+\gamma} $ can be obtained for the breathing and vertical mode eigenfrequencies of a slender torus orbiting a black hole \citep{blaes2006}, if the Newtonian limit is taken (in which the epicyclic frequencies are equal to one another, $\omega_r=\omega_\perp$). In general, just like the fundamental mode just considered, the vertical mode of the slender torus corresponds to rigid motion at the vertical epicyclic frequency, $\omega_1=\omega_\perp$ \citep{Kluzniak2005}, and the breathing mode of the levitating atmosphere is a direct analogue of the breathing mode of a slender torus.

\section{Radial oscillations in Newtonian framework}

We subject the static atmosphere to small radial perturbations of the form $\delta X(r,t) = \delta X_*(r)e^{i\omega t} $, where $\omega$ is the frequency with which perturbations propagate. Because of these small radial perturbations, fluid attains certain radial velocity given by $\delta v_r$. We then perform the linear perturbation analysis and derive the wave equation that governs the propagation of these perturbations. To do so, we assume that atmosphere is geometrically thin, i.e, $x \ll 1$. Relativistic corrections to energy density are negligible on such small scales to which these atmospheres extend. Therefore, Newtonian theory would suffice to evaluate $\omega$. In this work, we ignore the radiation drag, but note that Eq. (\ref{eqbrm}) is valid in any case as the  equilibrium atmosphere is static and the initial velocities being zero do not contribute to the pressure balance. 

We start with the equations of motion for a radially moving fluid in Newtonian framework. Linearized Euler and continuity equations are given as

	\begin{eqnarray}
	\dfrac{\partial}{\partial t}{\delta v_r}+ \dfrac{\partial}{\partial r}\left(\dfrac{\delta p}{\rho}\right)  &=& 0,\label{e23}\\
		\dfrac{\partial}{\partial t}(\delta \rho) + \dfrac{1}{r^2}\dfrac{\partial}{\partial r}\left(r^2 \rho \delta v_r\right)&=&0.\label{cont}
	\end{eqnarray}

We define a scalar potential $W= \delta p/(\omega \rho)$, which identifies with the fluid enthalpy, and simplifies our problem to great extent \citep{PP1984}. Considering the time derivative of the perturbation quantities would give us the following relations 
\begin{eqnarray}
\delta v_r &=& i \dfrac{\partial W}{\partial r}, \label{e26}\\
	\omega^ 2W  &=&i c_s^2 \left(\dfrac{\partial }{\partial r}{\delta v_r}+\dfrac{\delta v_r }{\rho}\dfrac{\partial \rho}{\partial r} + \dfrac{2 \delta v_r}{r}\right).
\label{e27}
\end{eqnarray}

Combining both of these equations, Eq. (\ref{e26}) and Eq. (\ref{e27}) and using Eq. (\ref{prho}), we obtain a second-order differential equation in $W$,
	\begin{eqnarray}
\dfrac{\partial ^2 W}{\partial r^2}+\left(\dfrac{n}{f}\dfrac{\partial f}{\partial r}+\dfrac{2}{r}\right)\dfrac{\partial W}{\partial r}+\dfrac{\omega^2}{c_s^2} W  &=& 0.
\end{eqnarray}

In general it is difficult to solve the above equation analytically but in the slender limit $(\beta \rightarrow 0)$ this equation simplifies to
\begin{eqnarray}
(1-\eta^2)\dfrac{\partial^2 W}{\partial \eta^2}-2n\eta \dfrac{\partial W}{\partial \eta}+ 2n \sigma^2 W  &=& 0 
\label{eigen}
\end{eqnarray}
where we have made use of the notations $\eta = x\omega_{r,0}/\beta$, with $f=1-\eta^2$, and $\sigma = \omega/\omega_{r,0}$. The eigenvalue problem given by Eq. (\ref{eigen}) has to be solved with appropriate boundary conditions. In this case, it is required that pressure and density vanish at the boundaries. Since we consider small, linear perturbations, this condition translates to vanishing of the Lagrangian perturbation in pressure at the boundaries in the form
\begin{eqnarray}
(\Delta p)_{f=0} = \left(\delta p +\xi \dfrac{\partial p}{\partial r} \right)_{f=0}=0,
\end{eqnarray}
where $\xi$ is the Lagrangian displacement. After rewriting the Lagrangian displacements in terms of $W$, the surface boundary condition becomes, 
\begin{eqnarray}
\left[\rho_0 f^n \omega_{r,0}\left( W \sigma-\dfrac{\eta}{\sigma}\dfrac{\partial W}{\partial \eta}\right)\right]_{f=0} = 0.
\end{eqnarray}
The requirement that $W$ and its first derivative be finite at $f=0$ is a sufficient condition for this boundary condition to hold. It is also possible that the singular eigenfunctions may appear on the surface boundaries. But such solutions must be rejected as they are not compatible with Eq. (\ref{eigen}). Therefore, the eigenfunctions are simply given by the Gegenbauer polynomials $C^{\mu}_{\nu}(\eta)$ with order $\mu = n-1/2$ and degree $\nu = 0,1,2,3...$, where $\nu(\nu+2n-1) = 2n \sigma^2$~\citep{abm1972}. Hence, 
\begin{eqnarray}
\omega_{\nu} = \sqrt{\dfrac{\nu(\nu+2n-1)}{2n}}\,\omega_{r,0}.
\label{freq}
\end{eqnarray}
with  $\omega_{r,0}^2 = (1-\lambda^2)^4c^6/(16G^2M^2 \lambda^4)$ by Eq.~(\ref{fundament}).

\section{Results and Discussions}

When the atmospheric fluid element in equilibrium is radially displaced, the restoring forces due the gravity and pressure variation act upon the fluid trying to restore it back to equilibrium. But owing to inertia, the displacement overshoots and thus the atmosphere is set into harmonic oscillations. Table~\ref{table1} summarizes the eigenfunctions and eigenfrequencies of the first few eigenmodes of this system. Degree $\nu = 0$ corresponds to a trivial solution with eigenfunction $1$, i.e., vanishing displacement by Eq. (\ref{e26}).
\begin{table*}
\caption{Eigenmodes and eigenfunctions of lowest order modes.}
\begin{tabular*}{ 2\columnwidth }{@{\extracolsep{\fill}} l c c c }
 \hline
~~~~~~~~$\nu$ & Eigenfrequency squared, $\sigma^2$ & Eigenfunction, $W$\\
 \hline
 \vspace{2mm}
~~~~~~~~0   & 0    &1\\
\vspace{2mm}
~~~~~~~~1&  1  & $(2n-1)\eta$ \\
\vspace{2mm}
~~~~~~~~2&$(2n+1)/n$& $\left(n-\dfrac{1}{2}\right)[\eta^2(2n+1)-1]$\\
\vspace{2mm}
~~~~~~~~3   &$3(n+1)/n$ & $\dfrac{\eta}{6}(4n^2-1)[(2n+3)\eta^2-3]$\\
\vspace{2mm}
~~~~~~~~4   &$2(2n+3)/n$ & $\dfrac{1}{24}(4n^2-1)[(2n+3)\eta^2((2n+5)\eta^2-6)+3]$\\
\hline
\label{table1}
\end{tabular*}
\end{table*}

The lowest order mode, $\nu=1$, is the fundamental mode given by the frequency $\omega_{\mathrm{f}} \equiv \omega_{1} = \omega_{r,0}$. In this mode the fluid oscillates around the ECS radially back and forth while maintaining the Lagrangian pressure profile in the atmosphere constant. Thus it is an incompressible mode. The displaced fluid is set into oscillations by the restoring force due to gravity.

For the next lowest order mode, $\nu=2$, pressure restoring forces become important. We term this as the breathing mode since the atmosphere expands and contracts symmetrically with respect to the ECS, with a frequency $ \omega_{\mathrm{b}}  \equiv \omega_{2} = \sqrt{(2n+1)/n}\,\omega_{r,0}$. 

\begin{figure}
\includegraphics[width=\columnwidth]{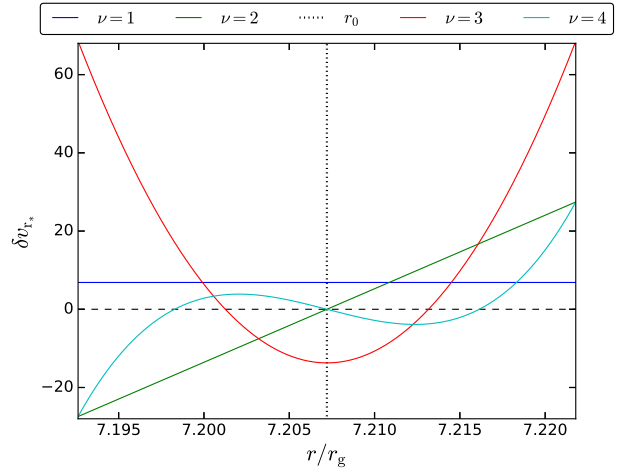}
\caption{Radial profiles of velocity perturbations for the lowest order modes. We take $n=3/2$ and $\lambda =0.85$. Horizontal dashed line is plotted to trace the nodes where the radial velocity goes zero, $\delta v_{\mathrm{r}} = 0$, while the vertical dotted line marks the location of the ECS. }
\label{mesh2}
\end{figure}

In Fig. \ref{mesh2}, we plot the radial profiles of the velocity perturbation. We consider $n=3/2$ and $\lambda =0.85$. We can scale the perturbation by an arbitrary value so that assumptions made to pursue linear perturbation analysis hold well. The domain of the plot corresponds to the extent of the equilibrium atmosphere. The vertical dotted line marks the location of the ECS, which for the given parameters is $\approx 7.207~r_{\mathrm{g}}$. The horizontal dashed line is plotted to identify the nodes at which velocity perturbations vanish. The fluid in the fundamental mode oscillates with uniform velocity depicted by the horizontal blue line. This means that at a given time, all the fluid particles in the atmosphere are moving with same speed in same direction. As the system evolves in this mode, the fluid oscillates back and forth through the horizontal dashed line. Velocity perturbation for the breathing mode holds a linear relation with radius as shown by the green line with a node at the ECS. While expanding and contracting, fluid in the atmosphere either recedes away from the ECS or approaches ECS, while the centre stays still marking a node at the ECS. As the system evolves, the green curve pivots around its node at the ECS. As we move further to the higher modes, the number of nodes simply increases.

\begin{figure}
\includegraphics[width=\columnwidth]{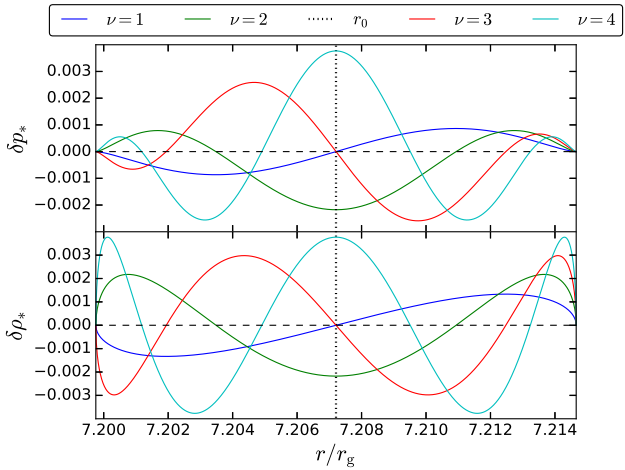}
\caption{Top panel shows the pressure eigenmode profile for lowest order modes ($\nu =1,~2,~3~{\rm and}~ 4$) calculated for $n=3/2$ and $\lambda =0.85$. Bottom panel shows the perturbed density variations for the same. Horizontal dashed line is where the perturbations vanish while the vertical dotted line marks the location of the ECS.}
\label{mesh1}
\end{figure}

In the last figure, Fig. \ref{mesh1}, the top and bottom panel show the radial profiles of the Eulerian perturbation in pressure and density respectively. The parameters considered here are same as before. The vertical dotted line carries the same meaning as discussed for the previous figure. The horizontal dashed line is plotted to identify the nodes at which the perturbations in density and pressure vanish. Note that the pressure and density perturbations vanish at the boundaries too as required by the boundary condition, but they are not to be counted as nodes. The fundamental mode plotted by the dark blue curve is an incompressible mode. But to see this from an Eulerian perturbation plot, we need the information of velocity perturbations. Since the velocity perturbation in the fundamental mode is uniform (see Fig. \ref{mesh1}), its divergence vanishes. This implies from the continuity equation that density or the pressure in a given volume remains the same.
For the breathing mode, the pressure drop at the centre is accompanied by an increase in pressure close to the boundaries as shown by the green curve with two nodes in the plot. Since we considered a polytropic fluid, the density perturbation profiles look analogous to the pressure perturbation profiles. Note that the number of nodes in pressure perturbations profile is one greater than the number of nodes in the velocity perturbations for a given mode.

From Table~\ref{table1}, the frequency of oscillations in the fundamental mode is simply $\omega_{r,0}$. This is independent of $n$ because the fundamental mode oscillations are caused purely by gravity. Remaining modes have the contribution from the fluid's pressure. The exact analytical expressions for eigenfrequencies allows us to calculate the ratio of the frequencies of the breathing mode, $\omega_{\mathrm{b}}$ to the fundamental mode, $\omega_{\mathrm{f}}$,
\begin{equation}
\dfrac{\omega_b}{\omega_f} = \sqrt{2+\dfrac{1}{n}}= \sqrt{1+\gamma}.
\label{ratio}
\end{equation}
For $n=3/2$, this ratio is $\sqrt{8/3} \approx 1.633$. Interestingly, the ratio of frequencies of any two modes is independent of $\lambda$, the stellar luminosity. 

For a neutron star with mass $1.4 M_{\odot}$ and $\lambda = 0.8$, we calculate the Newtonian frequency of oscillations for the $\nu=1$ fundamental mode to be  $1.171~\mathrm{kHz}$. Our result matches with the fundamental mode oscillation frequency earlier derived for full general relativistic equations~\citep{abarca2016}, if we consider the following relativistic effects. First, the invariant line element differs from coordinate displacement $\mathrm{d}r$, inducing a correction factor, $1/\sqrt{g_{rr}} \approx \lambda$. Secondly, in general relativity, proper time and coordinate time are different and the frequency measured by an observer at infinity would be redshifted by a factor $\sqrt{g_{tt}} \approx \lambda$. Owing to these two factors, if we multiply the Newtonian frequency by $\lambda^2$ we get the relativistic frequency $\tilde{\omega} = \dfrac{(1-\lambda^2)^2c}{4r_{\mathrm{g}}}$. For the above considered parameters, the relativistic frequency for the fundamental mode is then $\approx750~\mathrm{Hz}$, for $\lambda = 0.8$, which is the same as obtained in~\citet{abarca2016}. For $0.8<\lambda$ the fundamental frequency decreases in value with increasing luminosity, but similarly accounting for the relativistic corrections, the eigenfrequencies of the higher modes for $0.8<\lambda <0.98$ may fall in the observed range  of 300--600 Hz for X-ray burst quasi-periodic oscillations (QPOs). \citet{abarca2016} have also shown that for optically thin solutions, radiation drag can efficiently damp the fundamental mode oscillation. Here we find that the higher modes have larger frequencies than the damping rate found by those authors, so it is likely that the oscillations of the higher modes can be underdamped. We leave this for our future work, where we plan to study the full general relativistic motion of particles including radiation drag.

Also, note that $\omega_{r,0}$ increases with decreasing luminosity\footnote{For an impulsive change in luminosity the optically thin atmosphere may be ejected \citep{K2013,mishra2014}.}. So for any given mode, assuming that the polytropic index of the atmosphere remains the same, frequency of any of the eigenmodes would increase with time if the luminosity of the source decreases with time. This is similar to what is usually observed in the decay phase of the X-ray bursts \citep{stro2001,stro2006}.

\section{Conclusions}

In this paper, we provide a complete Newtonian analytic solution for the radial oscillation eigenmodes of a thin levitating atmospheric shell centred at the ECS around a luminous non-rotating stellar source. We derive the complete set of eigenfunctions for the modes, and the eigenfrequencies for all the mode oscillations\footnote{The atmospheric oscillations described here are very different from the oscillations of the ECS described in \citep{maciek2012}, which are due to accretion feedback and stellar luminosity variability.} are given by an exact analytical expression, Eq.~(\ref{freq}). We provide the ratio of the frequencies of breathing mode to the fundamental mode, $\sqrt{1+\gamma}$, which entirely depends on the polytropic index of the atmosphere. For $\gamma=5/3$, we find this ratio to be $\sqrt{8/3} \approx 1.633$, and for $\gamma=4/3$ it is $\sqrt{7/3} \approx 1.53$.

\section*{Acknowledgements}

We thank Maciek Wielgus and David Abarca for useful discussions throughout the work. This research was supported by the Polish NCN grants 2013/08/A/ST9/00795 and 2013/10/M/ST9/00729 and also in part by the National Science Foundation under Grant No. NSF PHY11-25915.

\bibliographystyle{mnras}
\bibliography{Newb_Newp}

\begin{thebibliography}{}
\makeatletter
\relax
\def\mn@urlcharsother{\let\do\@makeother \do\$\do\&\do\#\do\^\do\_\do\%\do\~}
\def\mn@doi{\begingroup\mn@urlcharsother \@ifnextchar [ {\mn@doi@}
  {\mn@doi@[]}}
\def\mn@doi@[#1]#2{\def\@tempa{#1}\ifx\@tempa\@empty \href
  {http://dx.doi.org/#2} {doi:#2}\else \href {http://dx.doi.org/#2} {#1}\fi
  \endgroup}
\def\mn@eprint#1#2{\mn@eprint@#1:#2::\@nil}
\def\mn@eprint@arXiv#1{\href {http://arxiv.org/abs/#1} {{\tt arXiv:#1}}}
\def\mn@eprint@dblp#1{\href {http://dblp.uni-trier.de/rec/bibtex/#1.xml}
  {dblp:#1}}
\def\mn@eprint@#1:#2:#3:#4\@nil{\def\@tempa {#1}\def\@tempb {#2}\def\@tempc
  {#3}\ifx \@tempc \@empty \let \@tempc \@tempb \let \@tempb \@tempa \fi \ifx
  \@tempb \@empty \def\@tempb {arXiv}\fi \@ifundefined
  {mn@eprint@\@tempb}{\@tempb:\@tempc}{\expandafter \expandafter \csname
  mn@eprint@\@tempb\endcsname \expandafter{\@tempc}}}

\bibitem[\protect\citeauthoryear{{Abarca} \& {Klu{\'z}niak}}{{Abarca} \&
  {Klu{\'z}niak}}{2016}]{abarca2016}
{Abarca} D.,  {Klu{\'z}niak} W.,  2016, \mn@doi [\mnras]
  {10.1093/mnras/stw1432}, \href
  {http://adsabs.harvard.edu/abs/2016MNRAS.461.3233A} {461, 3233}

\bibitem[\protect\citeauthoryear{{Abramowicz}, {Ellis}  \&
  {Lanza}}{{Abramowicz} et~al.}{1990}]{marek1990}
{Abramowicz} M.~A.,  {Ellis} G.~F.~R.,   {Lanza} A.,  1990, \mn@doi [\apj]
  {10.1086/169211}, \href {http://adsabs.harvard.edu/abs/1990ApJ...361..470A}
  {361, 470}

\bibitem[\protect\citeauthoryear{{Abramowitz} \& {Stegun}}{{Abramowitz} \&
  {Stegun}}{1972}]{abm1972}
{Abramowitz} M.,  {Stegun} I.~A.,  1972, {Handbook of Mathematical Functions}

\bibitem[\protect\citeauthoryear{{Bachetti} et~al.,}{{Bachetti}
  et~al.}{2014}]{bachetti}
{Bachetti} M.,  et~al., 2014, \mn@doi [\nat] {10.1038/nature13791}, \href
  {http://adsabs.harvard.edu/abs/2014Natur.514..202B} {514, 202}

\bibitem[\protect\citeauthoryear{{Basko} \& {Sunyaev}}{{Basko} \&
  {Sunyaev}}{1976}]{Basko1976}
{Basko} M.~M.,  {Sunyaev} R.~A.,  1976, \mn@doi [\mnras]
  {10.1093/mnras/175.2.395}, \href
  {http://adsabs.harvard.edu/abs/1976MNRAS.175..395B} {175, 395}

\bibitem[\protect\citeauthoryear{{Bini}, {Jantzen}  \& {Stella}}{{Bini}
  et~al.}{2009}]{bini2009}
{Bini} D.,  {Jantzen} R.~T.,   {Stella} L.,  2009, \mn@doi [Classical and
  Quantum Gravity] {10.1088/0264-9381/26/5/055009}, \href
  {http://adsabs.harvard.edu/abs/2009CQGra..26e5009B} {26, 055009}

\bibitem[\protect\citeauthoryear{{Blaes}, {Arras}  \& {Fragile}}{{Blaes}
  et~al.}{2006}]{blaes2006}
{Blaes} O.~M.,  {Arras} P.,   {Fragile} P.~C.,  2006, \mn@doi [\mnras]
  {10.1111/j.1365-2966.2006.10370.x}, \href
  {http://adsabs.harvard.edu/abs/2006MNRAS.369.1235B} {369, 1235}

\bibitem[\protect\citeauthoryear{{Coe}, {Burnell}, {Engel}, {Evans}  \&
  {Quenby}}{{Coe} et~al.}{1981}]{engel}
{Coe} M.~J.,  {Burnell} S.~J.~B.,  {Engel} A.~R.,  {Evans} A.~J.,   {Quenby}
  J.~J.,  1981, \mn@doi [\mnras] {10.1093/mnras/197.2.247}, \href
  {http://adsabs.harvard.edu/abs/1981MNRAS.197..247C} {197, 247}

\bibitem[\protect\citeauthoryear{{Fukue}}{{Fukue}}{1996}]{fukue1996}
{Fukue} J.,  1996, \mn@doi [\pasj] {10.1093/pasj/48.1.89}, \href
  {http://adsabs.harvard.edu/abs/1996PASJ...48...89F} {48, 89}

\bibitem[\protect\citeauthoryear{{Inogamov} \& {Sunyaev}}{{Inogamov} \&
  {Sunyaev}}{1999}]{is1999}
{Inogamov} N.~A.,  {Sunyaev} R.~A.,  1999, Astronomy Letters, \href
  {http://adsabs.harvard.edu/abs/1999AstL...25..269I} {25, 269}

\bibitem[\protect\citeauthoryear{{Israel} et~al.,}{{Israel}
  et~al.}{2016}]{israel2016}
{Israel} G.~L.,  et~al., 2016, preprint, \href
  {http://adsabs.harvard.edu/abs/2016arXiv160907375I} {} (\mn@eprint {arXiv}
  {1609.07375})

\bibitem[\protect\citeauthoryear{{Israel} et~al.,}{{Israel}
  et~al.}{2017}]{israel2017}
{Israel} G.~L.,  et~al., 2017, \mn@doi [\mnras] {10.1093/mnrasl/slw218}, \href
  {http://adsabs.harvard.edu/abs/2017MNRAS.466L..48I} {466, L48}

\bibitem[\protect\citeauthoryear{{Itanishi} \& {Fukue}}{{Itanishi} \&
  {Fukue}}{2017}]{2017PASJ...69...56I}
{Itanishi} Y.,  {Fukue} J.,  2017, \mn@doi [\pasj] {10.1093/pasj/psx033}, \href
  {http://adsabs.harvard.edu/abs/2017PASJ...69...56I} {69, 56}

\bibitem[\protect\citeauthoryear{{Kitabatake} \& {Fukue}}{{Kitabatake} \&
  {Fukue}}{2003}]{2003PASJ...55.1115K}
{Kitabatake} E.,  {Fukue} J.,  2003, \mn@doi [\pasj] {10.1093/pasj/55.6.1115},
  \href {http://adsabs.harvard.edu/abs/2003PASJ...55.1115K} {55, 1115}

\bibitem[\protect\citeauthoryear{{Klu{\'z}niak}}{{Klu{\'z}niak}}{2005}]{Kluzniak2005}
{Klu{\'z}niak} W.,  2005, \mn@doi [Astronomische Nachrichten]
  {10.1002/asna.200510420}, \href
  {http://adsabs.harvard.edu/abs/2005AN....326..820K} {326, 820}

\bibitem[\protect\citeauthoryear{{Klu{\'z}niak}}{{Klu{\'z}niak}}{2013}]{K2013}
{Klu{\'z}niak} W.,  2013, \mn@doi [\aap] {10.1051/0004-6361/201220150}, \href
  {http://adsabs.harvard.edu/abs/2013A%26A...551A..70K} {551, A70}

\bibitem[\protect\citeauthoryear{{Lewin}, {van Paradijs}  \& {van den
  Heuvel}}{{Lewin} et~al.}{1997}]{book26}
{Lewin} W.~H.~G.,  {van Paradijs} J.,   {van den Heuvel} E.~P.~J.,  1997,
  Cambridge Astrophysics Series, \href
  {http://adsabs.harvard.edu/abs/1997CAS....26.....L} {26}

\bibitem[\protect\citeauthoryear{{Mishra} \& {Klu{\'z}niak}}{{Mishra} \&
  {Klu{\'z}niak}}{2014}]{mishra2014}
{Mishra} B.,  {Klu{\'z}niak} W.,  2014, \mn@doi [\aap]
  {10.1051/0004-6361/201322188}, \href
  {http://adsabs.harvard.edu/abs/2014A%26A...566A..62M} {566, A62}

\bibitem[\protect\citeauthoryear{{Muno}, {Chakrabarty}, {Galloway}  \&
  {Psaltis}}{{Muno} et~al.}{2002}]{muno2002}
{Muno} M.~P.,  {Chakrabarty} D.,  {Galloway} D.~K.,   {Psaltis} D.,  2002,
  \mn@doi [\apj] {10.1086/343793}, \href
  {http://adsabs.harvard.edu/abs/2002ApJ...580.1048M} {580, 1048}

\bibitem[\protect\citeauthoryear{{Mushtukov}, {Suleimanov}, {Tsygankov}  \&
  {Ingram}}{{Mushtukov} et~al.}{2017}]{Mushtukov2017}
{Mushtukov} A.~A.,  {Suleimanov} V.~F.,  {Tsygankov} S.~S.,   {Ingram} A.,
  2017, \mn@doi [\mnras] {10.1093/mnras/stx141}, \href
  {http://adsabs.harvard.edu/abs/2017MNRAS.467.1202M} {467, 1202}

\bibitem[\protect\citeauthoryear{{Oh}, {Kim}  \& {Lee}}{{Oh}
  et~al.}{2010}]{oh2010}
{Oh} J.~S.,  {Kim} H.,   {Lee} H.~M.,  2010, \mn@doi [\prd]
  {10.1103/PhysRevD.81.084005}, \href
  {http://adsabs.harvard.edu/abs/2010PhRvD..81h4005O} {81, 084005}

\bibitem[\protect\citeauthoryear{{Papaloizou} \& {Pringle}}{{Papaloizou} \&
  {Pringle}}{1984}]{PP1984}
{Papaloizou} J.~C.~B.,  {Pringle} J.~E.,  1984, \mn@doi [\mnras]
  {10.1093/mnras/208.4.721}, \href
  {http://adsabs.harvard.edu/abs/1984MNRAS.208..721P} {208, 721}

\bibitem[\protect\citeauthoryear{{Rogers}}{{Rogers}}{2017}]{adam2017}
{Rogers} A.,  2017, \mn@doi [Universe] {10.3390/universe3010003}, \href
  {http://adsabs.harvard.edu/abs/2017Univ....3....3R} {3, 3}

\bibitem[\protect\citeauthoryear{{Sazonov}, {Sunyaev}  \& {Lund}}{{Sazonov}
  et~al.}{1997}]{groj}
{Sazonov} S.~Y.,  {Sunyaev} R.~A.,   {Lund} N.,  1997, in {Meyer-Hofmeister}
  E.,  {Spruit} H.,  eds,  Lecture Notes in Physics, Berlin Springer Verlag
  Vol. 487, Accretion Disks - New Aspects. p.~199, \mn@doi{10.1007/BFb0105833}

\bibitem[\protect\citeauthoryear{{Skinner}, {Bedford}, {Elsner}, {Leahy},
  {Weisskopf}  \& {Grindlay}}{{Skinner} et~al.}{1982}]{1982nat}
{Skinner} G.~K.,  {Bedford} D.~K.,  {Elsner} R.~F.,  {Leahy} D.,  {Weisskopf}
  M.~C.,   {Grindlay} J.,  1982, \mn@doi [\nat] {10.1038/297568a0}, \href
  {http://adsabs.harvard.edu/abs/1982Natur.297..568S} {297, 568}

\bibitem[\protect\citeauthoryear{{Stahl}, {Wielgus}, {Abramowicz},
  {Klu{\'z}niak}  \& {Yu}}{{Stahl} et~al.}{2012}]{stahl2012}
{Stahl} A.,  {Wielgus} M.,  {Abramowicz} M.,  {Klu{\'z}niak} W.,   {Yu} W.,
  2012, \mn@doi [\aap] {10.1051/0004-6361/201220187}, \href
  {http://adsabs.harvard.edu/abs/2012A%26A...546A..54S} {546, A54}

\bibitem[\protect\citeauthoryear{{Stahl}, {Klu{\'z}niak}, {Wielgus}  \&
  {Abramowicz}}{{Stahl} et~al.}{2013}]{stahl2013}
{Stahl} A.,  {Klu{\'z}niak} W.,  {Wielgus} M.,   {Abramowicz} M.,  2013,
  \mn@doi [\aap] {10.1051/0004-6361/201321595}, \href
  {http://adsabs.harvard.edu/abs/2013A%26A...555A.114S} {555, A114}

\bibitem[\protect\citeauthoryear{{Strohmayer}}{{Strohmayer}}{2001}]{stro2001}
{Strohmayer} T.~E.,  2001, \mn@doi [X-ray Astronomy: Stellar Endpoints, AGN,
  and the Diffuse X-ray Background] {10.1063/1.1434650}, \href
  {http://adsabs.harvard.edu/abs/2001AIPC..599..377S} {599, 377}

\bibitem[\protect\citeauthoryear{{Strohmayer} \& {Bildsten}}{{Strohmayer} \&
  {Bildsten}}{2006}]{stro2006}
{Strohmayer} T.,  {Bildsten} L.,  2006, {New views of thermonuclear bursts}.
pp 113--156

\bibitem[\protect\citeauthoryear{{Wielgus}}{{Wielgus}}{2016}]{maciekp}
{Wielgus} M.,  2016, in {Meiron} Y.,  {Li} S.,  {Liu} F.-K.,   {Spurzem} R.,
  eds,  IAU Symposium Vol. 312, Star Clusters and Black Holes in Galaxies
  across Cosmic Time. pp 131--134, \mn@doi{10.1017/S1743921315007693}

\bibitem[\protect\citeauthoryear{{Wielgus}, {Stahl}, {Abramowicz}  \&
  {Klu{\'z}niak}}{{Wielgus} et~al.}{2012}]{maciek2012}
{Wielgus} M.,  {Stahl} A.,  {Abramowicz} M.,   {Klu{\'z}niak} W.,  2012,
  \mn@doi [\aap] {10.1051/0004-6361/201220228}, \href
  {http://adsabs.harvard.edu/abs/2012A%26A...545A.123W} {545, A123}

\bibitem[\protect\citeauthoryear{{Wielgus}, {Klu{\'z}niak}, {Sa{\c d}owski},
  {Narayan}  \& {Abramowicz}}{{Wielgus} et~al.}{2015}]{maciek2015}
{Wielgus} M.,  {Klu{\'z}niak} W.,  {Sa{\c d}owski} A.,  {Narayan} R.,
  {Abramowicz} M.,  2015, \mn@doi [\mnras] {10.1093/mnras/stv2191}, \href
  {http://adsabs.harvard.edu/abs/2015MNRAS.454.3766W} {454, 3766}

\bibitem[\protect\citeauthoryear{{Wielgus}, {S{\c a}dowski}, {Klu{\'z}niak},
  {Abramowicz}  \& {Narayan}}{{Wielgus} et~al.}{2016}]{maciek2016}
{Wielgus} M.,  {S{\c a}dowski} A.,  {Klu{\'z}niak} W.,  {Abramowicz} M.,
  {Narayan} R.,  2016, \mn@doi [\mnras] {10.1093/mnras/stw548}, \href
  {http://adsabs.harvard.edu/abs/2016MNRAS.458.3420W} {458, 3420}

\bibitem[\protect\citeauthoryear{{van Straaten}, {van der Klis}, {Kuulkers}  \&
  {M{\'e}ndez}}{{van Straaten} et~al.}{2001}]{van2001}
{van Straaten} S.,  {van der Klis} M.,  {Kuulkers} E.,   {M{\'e}ndez} M.,
  2001, \mn@doi [\apj] {10.1086/320234}, \href
  {http://adsabs.harvard.edu/abs/2001ApJ...551..907V} {551, 907}

\makeatother
\end{thebibliography}

\label{lastpage}
\end{document}